\documentclass[12pt,preprint]{aastex}
\begin{document}

\title{Spin Evolution of Millisecond Magnetars with Hyperaccreting Fallback Disks: Implications
for Early Afterglows of Gamma-Ray Bursts}
\author{Z. G. Dai$^{1,2,*}$ and Ruo-Yu Liu$^{1,2,*}$}
\affil{$^{1}$School of Astronomy and Space Science, Nanjing University, Nanjing 210093, China \\
$^{2}$Key Laboratory of Modern Astronomy and Astrophysics (Nanjing
University), Ministry of Education, China \\
$^*${\rm dzg@nju.edu.cn, ryliu@nju.edu.cn}}

\begin{abstract}
The shallow decay phase or plateau phase of early afterglows of
gamma-ray bursts (GRBs), discovered by Swift, is currently
understood as being due to energy injection to a relativistic blast
wave. One natural scenario for energy injection invokes a
millisecond magnetar as the central engine of GRBs, because the
conventional model of a pulsar predicts a nearly constant
magnetic-dipole-radiation luminosity within the spin-down timescale.
However, we note that significant brightening occurs in some early
afterglows, which apparently conflicts with the above scenario. Here
we propose a new model to explain this significant brightening
phenomena by considering a hyperaccreting fallback disk around a
newborn millisecond magnetar. We show that for typical values of the
model parameters, sufficient angular momentum of the accreted matter
is transferred to the magnetar and spins it up. It is this spin-up
that leads to a dramatic increase of the magnetic dipole radiation
luminosity with time and thus significant brightening of an early
afterglow. Based on this model, we carry out numerical calculations
and fit well early afterglows of 12 GRBs assuming sufficiently
strong fallback accretion. If the accretion is very weak, our model
turns out to be the conventional energy-injection scenario of a
pulsar. Therefore, our model can provide a unified explanation for
the shallow decay phase, plateaus, and significant brightening of
early afterglows.
\end{abstract}


%%%%%%%%%%%%%%%%%%%%%%%%%%%%%%%%%%%%%%%%
\section {Introduction}
The successful launch of the Swift satellite (Gehrels et al. 2004)
has opened a new era of the study of cosmological gamma-ray bursts
(GRBs). In this era, there have been many important discoveries led
by Swift (Zhang 2007 and Gehrels et al. 2009 for recent reviews),
one of which is the identification of a canonical X-ray afterglow
light curve, described by broken power laws $F_\nu(t)\propto
t^{-\alpha}$ (Nousek et al. 2006; Zhang et al. 2006): an initial
steep decay phase with $\alpha\sim 3$ (or a steeper slope) extending
to $\sim 10^2-10^3$\,s is followed by a shallow decay phase with
$\alpha\sim 0.5$ or a flatter slope. This shallow decay phase
usually lasts $\sim 10^3-10^4$\,s. A subsequent normal decay phase
has $\alpha\sim 1.2$, being in agreement with the standard afterglow
model. No spectral evolution across the shallow-to-normal decay
break is observed. A post-jet-break decay phase with $\alpha\sim 2$,
predicted by the jet model, is sporadically observed following the
normal decay phase. Besides these four phases, one or multiple X-ray
flares also appear in nearly one half of GRB early afterglows. The
observed X-ray flares typically have very steep rising and decaying
slopes (Burrows et al. 2005; Falcone et al. 2007).

These observations suggest that the GRB central engine may be in a
long-lasting activity for two reasons. On one hand, the rapid rising
and decaying timescales and their distributions of X-ray flares
require that the central engine restarts at a later time (Lazzati \&
Perna 2007). This conclusion can also be drawn from the fact that
the peak time of the X-ray flare observed by Swift is nearly equal
to the ejection time of the outflow from the central engine, by
assuming that the decaying phase of an X-ray flare is due to the
high latitude emission from a relativistic outflow (Liang et al.
2006).

On the other hand, the shallow decay phase of early afterglows is
currently understood as being due to energy injection into a
relativistic blast wave, assuming an injection luminosity
$L(t)\propto t^{-q}$ (Zhang et al. 2006; Nousek et al. 2006). One
natural scenario invokes a strongly magnetic millisecond pulsar,
which spins down through magnetic dipole radiation (Dai \& Lu 1998a,
1998b; Zhang \& M\'esz\'aros 2001). An early version of this
scenario is that a newborn pulsar loses its rotational energy in the
form of Poynting flux. Numerical calculations based on this version
by Fan \& Xu (2006), Yu \& Huang (2009), and Dall'Osso et al. (2011)
show that it provides a satisfactory fitting to the observed shallow
decay phase. Furthermore, the observed lightcurve plateaus of the
early X-ray afterglows from GRBs 050319 (Huang et al. 2007), 050801
(De Pasquale et al. 2007), 060729 (Grupe et al. 2007), 070110 (Troja
et al. 2007), 080913 (Greiner et al. 2009), and 090515 (Rowlinson et
al. 2010) indicate that $q\simeq 0$, a pulsar-type energy injection.
A recent, more physical version of the energy injection scenario of
a pulsar assumes that the pulsar may continuously eject an
ultrarelativistic electron-positron-pair wind, interaction of which
with a circum-burst medium leads to a relativistic wind bubble (Dai
2004). The relativistic reverse shock emission from this bubble can
fit observed shallow decays and even plateaus in some GRB afterglows
(Yu \& Dai 2007; Mao et al. 2010).

An alternative scenario for energy injection requires ejecta with a
wide-$\Gamma$ (Lorentz factor) distribution from a newborn black
hole (Rees \& M\'esz\'aros 1998; Sari \& M\'esz\'aros 2000), in
which scenario the low-$\Gamma$ ejecta catch up with a blast wave
when the high-$\Gamma$ ejecta are decelerated. Two features of this
scenario are that all the materials with a wide-$\Gamma$
distribution are impulsively released from the central engine during
the prompt emission phase and that the resulting reverse shock
during the shallow decay phase is non-relativistic. On the contrary,
the central engine activity is long-lasting in the energy injection
scenario of a pulsar and the reverse shock is ultrarelativistic.
This difference has some astrophysical implications for testing the
two scenarios (Dai 2004; Yu et al. 2007; Corsi \& M\'esz\'aros
2009).

The comparisons of the energy injection scenario of a pulsar with
the observations (Lyons et al. 2010; Yu et al. 2010) show that the
central engine of some GRBs may be a millisecond magnetar, a type of
millisecond pulsar whose surface magnetic-field strength exceeds the
critical one. The conventional magnetic-dipole-radiation model
predicts $q\simeq 0$ within the spin-down timescale. This naturally
explains the shallow decay phase and plateaus ($\alpha\sim 0$). As
analyzed by Liang et al. (2007), however, we note that in some early
afterglows $\alpha$ is obviously smaller than zero. This requires
that $q<0$. In this paper we propose a new model to explain this
significant brightening phenomena by considering a hyperaccreting
fallback disk around a newborn millisecond magnetar. Zhang \& Dai
(2008, 2009) first investigated the properties of a hyperaccreting
disk around a neutron star, and Zhang \& Dai (2010) further studied
the effects of a magnetar-strength magnetic field on the disk. The
present paper focuses on the effects of a hyperaccreting fallback
disk on spin evolution of a newborn millisecond magnetar.

This paper is organized as follows: section 2 describes both the
generation of a GRB within the framework of a millisecond magnetar
and the magnetar-disk interactions. Section 3 studies the magnetar's
spin evolution during fallback accretion analytically and
numerically. We show that for typical values of the surface magnetic
field strength, initial rotation period, and accretion rate,
sufficient angular momentum of the accreted matter is transferred to
the magnetar and spins it up. It is this spin-up that leads to a
dramatic increase of the magnetic-dipole-radiation luminosity with
time and thus significant brightening of an early afterglow. Section
4 fits the early afterglows of 12 GRBs in the relativistic pulsar
wind bubble model proposed by Dai (2004) and reconsidered by Yu \&
Dai (2007). The final section presents a discussion and conclusions.

\section{GRB Generation and Magnetar-Disk Interactions}

\subsection{Generation of a GRB}
A millisecond magnetar has been one of the two leading models of
central engines of long-duration GRBs (Usov 1992; Duncan \& Thompson
1992; Thompson 1994; Wheeler et al. 2000; Thompson et al. 2004;
Woosley 2011 for a recent review). This model assumes that the
collapse and supernova explosion of a massive star at the end of its
life leaves behind a rapidly-rotating neutron star with a period of
$\sim 1\,$ms and an ultrastrong surface magnetic field of $\sim
10^{15}\,$G. An ultra-strong field may be produced by dynamo
processes (Duncan \& Thompson 1992; Akiyama et al. 2003; Thompson et
al. 2005). Subsequent cooling of the magnetar leads to a wind, which
has four phases (Thompson et al. 2004; Komissarov \& Barkov 2008;
Bucciantini et al. 2008, 2009; see Fig. 2 in Metzger et al. 2011 for
recent results): (1) in $\sim 1\,$s after the explosion, the wind is
driven by neutrino energy deposition, so it is matter-dominated. Its
asymptotic velocity is only $\sim 0.1c$. (2) A few seconds later,
the wind becomes magneto-centrifugally dominated. In this phase, it
is still matter-dominated and non-relativistic. (3) A few seconds
(perhaps $\sim 2-5\,$s) later after phase 2, the stellar mass-loss
rate decreases sufficiently so that the wind is accelerated to a
relativistic velocity possibly with Lorentz factor of $\geq 100$
from the magnetar surface out to the light cylinder by
magneto-centrifugal forces. It is the wind in phase 3 that may
generate a GRB. This wind is initially highly magnetized. Its
Lorentz factor evolves with radius as $\propto r^{1/3}$ and
eventually equals to the wind magnetization at the saturation radius
(Drenkhahn 2002; Drenkhahn \& Spruit 2002; also see Metzger et al.
2011). Beyond this radius, a series of internal collision-induced
magnetic reconnection and turbulence events would occur (Zhang \&
Yan 2011). These events could have a highly radiative efficiency for
producing a GRB. (4) At the end of the cooling epoch, both the
neutrino luminosity and mass-loss rate decrease dramatically. The
Lorentz factor of the wind increases to $\gamma_{\rm w}\sim 10^6$
and the millisecond magnetar enters its pulsar phase, in which its
rotational energy is lost through the magnetic dipole radiation
mechanism rather than the magneto-hydrodynamical mass-loss process.

A newborn millisecond magnetar cools on the Kelvin-Helmholtz
timescale ($t_{\rm KH}$) by radiating its gravitational binding
energy via neutrinos. Thus, the duration of a GRB powered by the
magnetar is nearly equal to $t_{\rm KH}$. For a non-rotating
proto-neutron star, $t_{\rm KH}$ is about 30\,s (Pons et al. 1999).
For the millisecond magnetar, however, rapid rotation may decrease
the overall neutrino luminosity and average energy by a factor of at
most $\sim 5-6$ (Thompson et al. 2005). If the same amount of
gravitational binding energy is assumed to be liberated under the
zeroth-order approximation, we expect that $t_{\rm KH}$ should
increase by the same factor. In the millisecond magnetar model,
therefore, the maximum duration of a GRB is unlikely to be longer
than $\sim 200\,$s.

\subsection{Magnetar-Disk Interactions}
The materials ejected during the supernova explosion must have a
velocity distribution, some of which fail to achieve escape velocity
and eventually fall back onto the central millisecond magnetar. The
minimum free-fall time of fallback matter is denoted by $t_{\rm
fb}$. This time might be extended because the falling matter needs
to overcome the resistance of a low-density neutrino-heated bubble
(MacFadyen et al. 2001). If this effect is neglected, $t_{\rm fb}$
corresponds to the minimum radius around which matter starts to fall
back, $r_{\rm fb}=(2GMt_{\rm fb}^2)^{1/3}\simeq 1.0\times
10^{10}(M/1.4M_\odot)^{1/3}(t_{\rm fb}/50\,{\rm s})^{2/3}\,{\rm
cm}$, where $M$ is the magnetar mass. However, the matter staying
around $r_{\rm fb}$ cannot be immediately accreted. This is because
a relativistic wind (with luminosity of $L_{\rm w}\sim
10^{49}-10^{51}\,{\rm erg}\,{\rm s}^{-1}$) from the central magnetar
may exert an outward ram pressure, which stops fallback accretion.
Once $L_{\rm w}$ decreases dramatically just at $t_{\rm KH}$,
fallback accretion may be able to proceed. Therefore, we obtain the
time when fallback accretion is expected to start, $t_0=\max(t_{\rm
fb},t_{\rm KH})\sim 10^2\,$s. This time is similar to the one from
numerical simulations of MacFadyen et al. (2001). For simplicity, we
set $t_0=100\,$s in the following calculations.

Following MacFadyen et al. (2001) and Zhang et al. (2008), we
parameterize the fallback accretion rate
\begin{equation}
\dot{M}=(\dot{M}_{\rm early}^{-1}+\dot{M}_{\rm late}^{-1})^{-1},
\end{equation}
where
\begin{equation}
\dot{M}_{\rm early}=10^{-3}\eta t^{1/2}M_\odot\,{\rm s}^{-1},
\end{equation}
and
\begin{equation}
\dot{M}_{\rm late}=10^{-3}\eta t_1^{13/6}t^{-5/3}M_\odot\,{\rm
s}^{-1}.
\end{equation}
Here $\eta\sim 0.01-10$ is a factor that accounts for different
explosion energies (smaller $\eta$ corresponds to a more energetic
explosion), and $t_1\sim 200-10^3\,$s is the time at which the mass
accretion rate starts to drop (longer $t_1$ for smaller $\eta$). For
$t\gg t_1$, equation (1) shows the late-time fallback accretion
behavior follows $\dot{M}\propto t^{-5/3}$, as suggested by
Chevalier (1989). Assuming that $M_0$ is the initial baryonic mass
of the magnetar, from equation (1), we obtain the stellar total
baryonic mass at time $t$,
\begin{equation}
M_b(t)=M_0+\int_0^t\dot{M}dt.
\end{equation}
Because a non-negligible fraction of this mass becomes binding
energy and is radiated away in the form of neutrinos (as discussed
by Lattimer \& Prakash 2001), the time-dependent gravitational mass
of the accreting magnetar with radius $R_s$ is
\begin{equation}
M=M_b(t)\left[1+\frac{3}{5}\frac{GM_b(t)}{R_sc^2}\right]^{-1}.
\end{equation}

Since fallback matter has sufficient angular momentum, a
geometrically-thin hyperaccreting disk forms around a magnetar,
similar to the black hole disk of Chen \& Beloborodov (2007). One
main difference between these two types of disk is that the black
hole disk extends to the innermost stable orbit but the magnetar
disk is truncated at some radius by the magnetic field. This inner
termination radius is one of the most pivotal physical quantities
for the magnetar-disk interactions, as it affects the flow of energy
and angular momentum in the accretion process. According to the
popular viewpoint (Davidson \& Ostriker 1973; Illarinov \& Sunyaev
1975), since it is fixed around the magnetospheric radius, the inner
termination radius increases smoothly across the corotation radius
as the mass accretion rate decreases. Thus, once the inner
termination radius is beyond the corotation radius, matter may be
ejected from the system by the super-Keplerian magnetosphere, that
is, the disk is in the propeller regime. This viewpoint was recently
adopted to explore the spin evolution of a newborn millisecond
magnetar and propose a propeller-powered supernova as a new
mechanism for supernovae (Piro \& Ott 2011).

We next define three useful radii within the accretion disk. The
first radius is the corotation radius at which the Keplerian angular
velocity ($\Omega_{\rm K}$) is equal to the rotation angular
velocity of the central magnetar ($\Omega_s$),
\begin{equation}
r_c=\left(\frac{GM}{\Omega_s^2}\right)^{1/3}.
\end{equation}
The second radius is the magnetospheric radius defined by
\begin{equation}
r_m=\left(\frac{\mu^4}{GM\dot{M}^2}\right)^{1/7},
\end{equation}
where $\mu=B_0R_s^3$ is the magnetic dipole moment of the magnetar
and $B_0$ is the surface magnetic field. The third radius is the
distance from the stellar center to the light cylinder,
\begin{equation}
R_{\rm L}=\frac{c}{\Omega_s}.
\end{equation}
Within this radius, the vertical magnetic field component of the
disk is assumed to have the dipolar form, $B_z=\mu/r^3$. In
addition, the fastness parameter is defined as the ratio of the
stellar rotation frequency to the Keplerian angular velocity at the
magnetospheric radius,
\begin{equation}
\omega=\frac{\Omega_s}{\Omega_{\rm
K}(r_m)}=\left(\frac{r_m}{r_c}\right)^{3/2}.
\end{equation}

The accretion disk is assumed to be truncated at $r=r_m$. If
$r_m<r_c$, the matter at $r_m$ is accreted onto the magnetar along
some magnetic field lines and forced to corotate with the magnetar,
so that angular momentum of the accreted matter is always
transferred to the magnetar. This provides a positive torque for the
magnetar, $\tau_0=\dot{M}\sqrt{GMr_m}$. If $r_m\geq r_c$ (viz., in
the propeller phase), however, the accreted matter at $r_m$
initially rotates at the Keplerian angular frequency and immediately
at the stellar angular velocity by the ``magnetic slingshot"
mechanism. This propeller effect thus leads to a negative torque
exerted on the magnetar, $\tau_0=\dot{M}\sqrt{GMr_m}(1-\omega)$.

For $r>r_m$, the differential motion between the Keplerian disk and
the magnetar generates an azimuthal field component, $B_\phi$. Wang
(1995) derived some different expressions for $B_\phi$ that are
dependent on the field dissipation mechanisms. Here we adopt a
simple but physically plausible expression for $B_\phi$ as a
function of radial distance (Livio \& Pringle 1992; Wang 1995;
Rappaport et al. 2004; Klu\'zniak \& Rappaport 2007),
\begin{equation}
B_\phi=B_z\left(1-\frac{\Omega_{\rm
K}}{\Omega_s}\right)=\left(\frac{\mu}{r^3}\right)\left[1-\left(\frac{r_c}{r}\right)^{3/2}\right].
\end{equation}
The magnetic torque exerted on the magnetar by the disk is given by
\begin{eqnarray}
\tau_{\rm M} & = & -\int_{r_m}^{R_{\rm L}}r^2B_\phi B_zdr\nonumber\\
& = & -\int_{r_m}^{R_{\rm
L}}\frac{\mu^2}{r^4}\left[1-\left(\frac{r_c}
{r}\right)^{3/2}\right]dr\nonumber\\
& = & -\frac{\mu^2}{9}\left(\frac{3}{r_m^3}-\frac{3}{R_{\rm
L}^3}-2\sqrt{\frac{r_c^3}{r_m^9}}+2\sqrt{\frac{r_c^3}{R_{\rm
L}^9}}\right)\nonumber\\ & = & -\frac{\mu^2}{9r_m^3}
\left(3-3\epsilon^2-\frac{2}{\omega}+\frac{2\epsilon}{\omega}\right),
\end{eqnarray}
where $\epsilon=(r_m/R_{\rm L})^{3/2}$. Therefore, the net torque
exerted on the magnetar by the accretion disk reads
\begin{equation}
\tau_{\rm acc}=\tau_0+\tau_{\rm
M}=n(\epsilon,\omega)(\dot{M}\sqrt{GMr_m})=n(\epsilon,\omega)\frac{\mu^2}{r_m^3},
\end{equation}
where $n(\epsilon,\omega)$ is the dimensionless torque parameter,
\begin{equation}
n(\epsilon,\omega)=\left\{\begin{array}{ll}(2-2\epsilon+6\omega+3\epsilon^2\omega)/(9\omega),
& {\rm for}\,\,\omega<1,
\\ (2-2\epsilon+6\omega+3\epsilon^2\omega-9\omega^2)/(9\omega),
& {\rm for}\,\,\omega\geq 1.\end{array} \right.
\end{equation}
Please note that neither $\tau_0$ nor $n(\epsilon,\omega)$ connects
smoothly from $\omega<1$ to $\omega>1$ in our model. This is because
for $\omega<1$ and $\omega>1$, the system is in two different
phases, the accretion phase and the propeller phase. Recently a
discontinuity of the dimensionless torque parameter across
$\omega=1$ was also noted by Tauris (2012). When $r_m>R_{\rm L}$
(viz., $\epsilon>1$), the magnetar-disk interactions are so weak
that $\tau_{\rm acc}=0$. In this case, the magnetar behaves as a
normal pulsar, which spins down through magnetic dipole radiation.

\section{Spin Evolution of the Magnetar}
Spin evolution of the magnetar is given by the following
differential equation,
\begin{equation}
\frac{d(I\Omega_s)}{dt}=\tau_{\rm acc}+\tau_{\rm dip},
\end{equation}
where $I=0.35MR_s^2$ is the stellar moment of inertia and $\tau_{\rm
dip}$ is the torque due to magnetic dipole radiation,
\begin{equation}
\tau_{\rm
dip}=-\frac{\mu^2\Omega_s^3\sin^2\chi}{6c^3}=-\frac{\mu^2\sin^2\chi}{6R_{\rm
L}^3},
\end{equation}
with $\chi$ being the inclination angle of the magnetic axis to the
rotation axis. For moderately stiff to stiff equations of state for
nuclear matter, the radius of a massive neutron star is nearly
independent on the mass (Lattimer \& Prakash 2001), so $R_s$ is
taken to be a constant in this paper.

\subsection{Asymptotic Analytical Solutions}
Before carrying out numerical calculations on equation (14), we
derive analytical solutions in three limiting cases. Generally, we
have $r_m \ll R_L$ or $\epsilon \ll 1$, so $n(\epsilon,\omega)$ is
simplified as
\begin{equation}
n(\epsilon,\omega)\simeq \frac{2}{9\omega}+\frac{2}{3} >
\frac{8}{9},\,\,\,{\rm for} ~\omega <1,
\end{equation}
and
\begin{equation}
n(\epsilon,\omega)\simeq\frac{3-(3\omega-1)^2}{9\omega} \leq
-\frac{1}{9},\,\,\,{\rm for} ~\omega \geq 1.
\end{equation}
Hence, we obtain
\begin{equation}
\left|\frac{\tau_{\rm dip}}{\tau_{\rm acc}}\right|=
\frac{\epsilon^2{\rm sin}^2\chi}{|6n(\epsilon,\omega)|}\ll 1,
\end{equation}
and $\tau_{\rm dip}$ can be neglected compared to $\tau_{\rm acc}$
in equation (14). Then first, for $\omega\ll 1$ (viz., slow
rotators), we find $n(\epsilon,\omega)\simeq 2/(9\omega)$ from
equation (13). After further assuming a constant moment of inertia
and neglecting the $\dot{I}\Omega_s$ term in equation (14), we have
\begin{equation}
I\frac{d\Omega_s}{dt}\simeq \tau_{\rm acc}\propto
\dot{M}^{9/7}\Omega_s^{-1}.
\end{equation}
Since $\dot{M}\propto t^{1/2}$ at early times ($t_0<t<t_1$),
equation (19) becomes
\begin{equation}
\Omega_s\propto t^{23/28},
\end{equation}
showing that the magnetar spins up at early times.

Second, in the case of $\omega\sim 1$ , we obtain
$\Omega_s\sim\Omega_{\rm K}(r_m)\propto r_m^{-3/2}\propto
\dot{M}^{3/7}$. Since $\dot{M}\propto t^{-5/3}$ at late times, we
find
\begin{equation}
\Omega_s\propto t^{-5/7},
\end{equation}
which is roughly consistent with an approximative solution to
equation (14), $\Omega_s\propto t^{-3/7}$, if we also assume a
constant moment of inertia and neglect the $\dot{I}\Omega_s$ and
$\tau_{\rm dip}$ terms as in the first case. This shows that the
magnetar spins down.

Third, if $\omega\gg 1$ (viz., fast rotators), we find
$n(\epsilon,\omega)\simeq -\omega$. From equation (14) together with
$\dot{M}\propto t^{-5/3}$, we have
\begin{equation}
\frac{d\Omega_s}{dt}=-K\Omega_st^{-5/7},
\end{equation}
assuming that $K$ is the coefficient. An integration to equation
(22) leads to
\begin{equation}
\Omega_s=\Omega_s(t_1)\exp[-(7K/2)(t^{2/7}-t_1^{2/7})].
\end{equation}
This also implies that the magnetar spins down.

\subsection{Numerical Results}
Numerical integrations to equation (14) lead to spin evolution of
the magnetar for different values of the model parameters (viz.,
surface magnetic field strength $B_0$, initial rotation period
$P_0$, $\eta$, and $t_1$). This further provides the
magnetic-dipole-radiation luminosity as a function of
time\footnote{The coefficient in this equation (Shapiro \& Teukolsky
1983) is by a factor of 4 smaller than that adopted in Dai \& Lu
(1998a).},
\begin{equation}
L_{\rm dip}=\frac{\mu^2\Omega_s^4\sin^2\chi}{6c^3}=9.6\times
10^{48}\,{\rm erg}\,{\rm s}^{-1}\sin^2\chi
\left(\frac{\mu}{10^{33}\,{\rm G}\,{\rm
cm}^3}\right)^2\left(\frac{P}{1\,{\rm ms}}\right)^{-4}.
\end{equation}

In our calculations, we take $\sin^2\chi=0.5$. For this value and
smaller values of $\sin^2\chi$ (viz., weakly oblique rotators), the
magnetar-disk interaction and dimensionless torque parameter are
nearly identical to those for an aligned rotator (Wang 1997). The
initial baryonic mass of a magnetar is assumed to be $1.4M_\odot$
and the maximum gravitational mass is $2.5M_\odot$, beyond which the
magnetar may become a black hole. We consider this value of the
maximum mass of a neutron star for two reasons: (1) very stiff
nuclear equations of state lead to the maximum mass of $\sim
2.5M_\odot$ (Lattimer \& Prakash 2001), and more importantly, (2)
detections of the mass of the black widow pulsar, PSR B1957+20, give
$M_{\rm PSR}=(2.40\pm0.12)M_\odot$ (van Kerkwijk et al. 2011). In
addition, a postmerger millisecond pulsar with mass $\geq
2.5M_\odot$ has been suggested by Dai et al. (2006) to explain X-ray
flares from a short-duration GRB. Figure 1 shows evolution of the
stellar mass with time. We see that for typical values of $\eta$ and
$t_1$ the stellar mass does not exceed the maximum mass during the
fallback accretion.

We take the benchmark model parameters: $P_0=3\,$ms,
$B_0=10^{15}\,$G, $\eta=0.5$, and $t_1=400\,$s. Figure 2 plots the
stellar rotation period as a function of time for these benchmark
values. We can see (from the red line in this figure) that the
magnetar first spins up and then spins down. The evolutional
behaviors of the magnetar's spin at early and late times are
consistent with the asymptotic analytical solutions given by
equations (20) and (21) respectively. The results for some other
values of the model parameters are also shown in Figure 2. We also
see that at late times the magnetar always spins down, being
independent of what values the model parameters are taken to be.
This is due to the fact that the fastness parameter $\omega$ is
close to (and somewhat larger than) unity at late times. At early
times, however, the longer initial rotation period (or weaker
surface magnetic field strength or larger $\eta$ or longer $t_1$),
the more significant initial spin-up. Figure 3 shows the
magnetic-dipole-radiation luminosity ($L_{\rm dip}$) as a function
of time. We find a dramatic increase of $L_{\rm dip}$ for typical
values of the model parameters.

We also calculate the magnetar's rotation parameter $\beta=T/|W|$,
where $T=I\Omega_s^2/2$ and $|W|$ is given by (Lattimer \& Prakash
2001)
\begin{equation}
|W|\simeq 0.6Mc^2\frac{GM/R_sc^2}{1-0.5(GM/R_sc^2)},
\end{equation}
which is shown in Figure 4. We see $\beta<0.14$ for typical values
of the model parameters (except for the blue line in the right-upper
panel), implying that some instabilities such as dynamical bar-mode
instabilities and secular instabilities can be neglected. This is
because the occurrence of these instabilities requires $\beta>0.27$
(Chandrasekhar 1969) and $\beta>0.14$ (Lai \& Shapiro 1995)
respectively.

\section{Fitting to Early Afterglows}
The interaction of an ultrarelativistic wind from the millisecond
magnetar with its ambient medium is in physics similar to the
well-observed Crab Nebula. In order to explain the Crab Nebula, it
was proposed (Rees \& Gunn 1974; Kennel \& Coroniti 1984; Begelman
\& Li 1992; Chevalier 2000) that a realistic, continuous wind from
the Crab pulsar is ultrarelativistic and dominated by the kinetic
energy flux of electron-positron pairs. From the viewpoint of
evolution, even if this wind is initially Poynting flux-dominated,
the fluctuating component of the magnetic field in the wind can be
dissipated by magnetic reconnection and used to accelerate the wind
to an ultrarelativistic velocity (Coroniti 1990; Michel 1994; Kirk
\& Sk{\ae}jaasen 2003). Recently, Aharonian et al. (2012) suggested
that the acceleration should take place abruptly in the narrow
cylindrical zone with radius between $20R_{\rm L}$ and $50R_{\rm L}$
and the wind's Lorentz factor $\gamma_{\rm w}\sim 10^6$ to fit the
spectral energy distribution of the pulsed high-energy $\gamma$-ray
radiation from the Crab pulsar, even though this suggestion
challenges current models on wind acceleration. In the case of a GRB
afterglow, therefore, if the central engine is a millisecond
magnetar, we assume that the magnetar's wind with luminosity of
$L_{\rm w}\simeq L_{\rm dip}$ is accelerated to a Lorentz factor of
$\gamma_{\rm w}\sim 10^6$ within a cylinder of radius much less than
the typical deceleration radius ($\sim 10^{16}-10^{17}\,$cm) of a
relativistic GRB fireball in an interstellar medium. Please note
that this assumption relaxes the requirement of abrupt acceleration
of an ultrarelativistic wind suggested by Aharonian et al. (2012),
but still keeps $\gamma_{\rm w}\sim 10^6$. A similar value of
$\gamma_{\rm w}$ has been adopted for some pulsar wind nebulae,
e.g., G0.9+0.1 (Tanaka \& Takahara 2011), and required by Suzaku
observations of PSR B1259-63 (Uchiyama et al. 2009). As we find in
our calculations, a large value of $\gamma_{\rm w}$ favors the
occurrence of a lightcurve plateau or brightening of an early
afterglow, although an accurate value of $\gamma_{\rm w}$ remains
highly uncertain in the literature.

The interaction of an ultrarelativistic wind with its ambient medium
leads to a relativistic wind bubble (Dai 2004; Yu \& Dai 2007). This
can be regarded as a relativistic version of the Crab Nebula. The
relativistic wind bubble should include two shocks: a reverse shock
that propagates into the cold wind and a forward shock that
propagates into the ambient medium. Thus, there are four regions
separated in the bubble by these shocks: (1) the unshocked medium,
(2) the forward-shocked medium, (3) the reverse-shocked wind gas,
and (4) the unshocked cold wind, where regions 2 and 3 are separated
by a contact discontinuity. Dai (2004) analyzed the wind bubble's
dynamics and emission features, and found a plateau of the reverse
shock emission light curve. Yu \& Dai (2007) and Mao et al al.
(2010) carried out numerical calculations and confirmed such a
plateau feature for typical values of the model parameters. This
feature is due to the fact that for a magnetar without any accretion
the wind luminosity $L_{\rm w}$ is nearly a constant at early times
less than the typical spin-down timescale. As in sections 2 and 3,
the fallback accretion spins up the magnetar, leading to an increase
of the wind luminosity with time at early times, for typical values
of the model parameters. It is thus expected that the reverse
emission gives rise to significant brightening of an early
afterglow.

Following Yu \& Dai (2007), we calculate the dynamics of a
relativistic wind bubble expanding in an interstellar medium (ISM)
and the emission fluxes of forward and reverse shocks. As in Sari et
al. (1998), we assume that $p$ is the spectral index of the
shock-accelerated electrons, and the electron and magnetic energy
densities behind a shock are fractions, $\epsilon_e$ and
$\epsilon_B$, of the total energy density of the shocked matter
respectively. Of course, these parameters may be different for
forward and reverse shocks, as the unshocked medium and the
unshocked wind may have different magnetic fields and compositions.
Figure 5 shows the light curves of forward and reverse shock
emissions for the benchmark values of the model parameters (i.e.,
$P_0=3\,$ms, $B_0=10^{15}\,$G, $\eta=0.5$, and $t_1=400\,$s). From
this figure, we can see significant brightening of an early
afterglow, being due to a dramatic increase of the magnetar's wind
luminosity with time.

We search GRBs detected by
Swift\footnote{http://www.swift.ac.uk/xrt\_curves/} and find the
early-time significant brightening of 12 afterglows. Figure 6
provides fitting to these early afterglows for the assumed initial
rotation period, surface magnetic field, and the parameters involved
in the accretion rate. The required shock parameters are shown in
Table 1. This table also presents the stellar gravitational mass at
$10^6\,$s [i.e., $M(10^6\,{\rm s})$], the maximum value of the
rotation parameter ($\beta_{\rm max}$), and the minimum
magnetospheric radius ($r_{m,{\rm min}}$) for each of 12 GRBs. We
can see $M(10^6\,{\rm s})<2.5M_\odot$, $\beta_{\rm max}<0.14$, and
$r_{m,{\rm min}}>R_s$. In addition, $R_{\rm
L}=c/\Omega_s=47.7(P/1\,{\rm ms})\,$km is much greater than $R_s$.
These ensure that our model is self-consistent. Figure 6 together
with Table 1 shows that our model can well explain the significant
brightening of 12 early afterglows. This explanation requires that
the fallback accretion is sufficiently strong.

When the fallback accretion rate is so small that the magnetospheric
radius $r_m$ exceeds the light-cylinder radius $R_{\rm L}$ (viz.,
$\epsilon>1$), both the fallback accretion and the propeller effect
stop and the torque exerted on the magnetar by the accretion disk
disappears. Meanwhile, the magnetar spins down only via the magnetic
dipole radiation mechanism, which has been shown to be able to
explain the shallow decay phase or the plateau phase of early
afterglows (Dai \& Lu 1998a, 1998b; Zhang \& M\'esz\'aros 2001; Dai
2004; Yu \& Dai 2007; Mao et al. 2010). In this case, therefore, our
present model turns out to be the conventional energy injection
scenario of a pulsar.

\section{Discussion and Conclusions}
Several physical explanations of the shallow decay phase or the
plateau phase of early afterglows discovered by Swift include energy
injection invoking a long-lasting central engine, energy injection
from ejecta with a wide-$\Gamma$ distribution, two-component jets,
dust scattering, varying microphysical parameters, and so on (see
Zhang 2007). The two leading scenarios are based on energy injection
to a relativistic blast wave. The first scenario invokes a
millisecond magnetar, while the second scenario in fact requires a
stellar-mass black hole. Thus, these two scenarios have different
central engines. Three astrophysical implications are discussed to
test them.

First, in the first scenario, the magnetic field in the
reverse-shocked region of a relativistic wind bubble consists of two
components: a large-scale toroidal field and a random field. If the
toroidal field dominates over the random component, one would expect
high polarization of an early afterglow during the shallow decay
phase, as discussed by Dai (2004). This could be used to distinguish
between the relativistic-wind-bubble model and the other
explanations including the second scenario.

Second, the compositions of winds in the two scenarios are also
different: the wind is lepton-dominated in the first scenario and
baryon-dominated in the second scenario. Yu et al. (2007) studied
the dynamics of winds and calculated the corresponding high-energy
photon emission by considering synchrotron radiation and inverse
Compton scattering of electrons. Even though in the two scenarios
there is a plateau (or a bump) in high-energy light curves during
the X-ray shallow decay phase, the first scenario predicts more
significant high-energy gamma-ray afterglow emission than the second
scenario does. This is because a considerable fraction of the
injecting energy in the first scenario is shared by a relativistic
long-lasting reverse shock and the reverse-shock energy is almost
carried by leptons (electrons and positrons), while the energy of a
nonrelativistic reverse shock is mainly carried by baryons in the
second scenario.

Third, if the initial period of a newborn magnetar is as small as
$\sim 1\,$ms and even smaller, then the stellar rotation parameter
$\beta$ may be larger than $0.14$ (e.g., the blue line in the
right-upper panel of Figure 4). In this case, some instabilities
such as dynamical bar-mode instabilities and secular instabilities
could not only be motivated so that they affect the stellar spin
evolution, but also the resultant gravitational waves would be
detectable with the future advanced-LIGO detector for a nearby GRB.
Meanwhile, the magetar's rotational energy could be injected to a
post-burst blast wave via magnetic dipole radiation, leading to the
shallow decay phase of an early afterglow (Corsi \& M\'esz\'aros
2009).

To summarize. It is well known that the conventional model of a
pulsar predicts a nearly constant magnetic-dipole-radiation
luminosity within the spin-down timescale. This provides an
explanation for the shallow decay phase or the plateau phase of
early afterglows. However, we note that significant brightening
occurs in some early afterglows, which apparently conflicts with the
conventional model. In order to explain this significant brightening
phenomena, we here investigate the effect of a hyperaccreting
fallback disk on the spin evolution of a newborn millisecond
magnetar. We show that for typical values of the model parameters,
sufficient angular momentum of the accreted matter is transferred to
the magnetar and spins it up. It is this spin-up that leads to a
dramatic increase of the magnetic-dipole-radiation luminosity with
time and thus significant brightening of an early afterglow.
Furthermore, we carry out numerical calculations and fitted well
early afterglows of 12 GRBs assuming sufficiently strong fallback
accretion. Furthermore, It is worth noting that if the accretion is
very weak, our present model turns out to be the previously-proposed
energy injection scenario of a pulsar. Therefore, our model can
provide a unified explanation for the shallow decay phase, plateaus,
and significant brightening of early afterglows. In addition,
possible detections of high polarization, gravitational waves,
and/or high-energy gamma-rays during the shallow decay phase would
be used to test this model in the future.

\acknowledgments We thank the referee for useful comments and X.-D.
Li and S. Rappaport for helpful discussions. This work was supported
by the National Natural Science Foundation of China (grant no.
11033002).

\begin{table}
\begin{center}
\caption{Shock parameters and some other model parameters for
fitting light curves of X-ray afterglows of some GRBs. The ISM
number density, the magnetic field equipartition factor in the
forward shock, the electron equipartition factor in the reverse
shock, the spectral power-law index of forward-shocked electrons,
the initial bulk Lorentz factor of the fireball and the bulk Lorentz
factor of the relativistic wind are taken to be $n=1\,{\rm
cm}^{-3}$, $\epsilon_{B,f}=0.1$, $\epsilon_{e,r}=1-\epsilon_{B,r}$,
$p_f=2.2$, $\gamma_0=300$, and $\gamma_{\rm w}=10^6$ for all GRBs
(where the subscripts $f$ and $r$ denote forward and reverse shocks
respectively). In this table, $E_{{\rm iso},51}$ is the postburst
initial fireball energy in units of $10^{51}$ ergs, $M(10^6\,{\rm
s})$ is the mass of the accreting magnetar at $10^6$ s, $\beta_{\rm
max}$ is the maximum value of the magnetar's rotation parameter, and
$r_{m,{\rm min}}$ is the minimum value of the magnetospheric radius.
The redshifts of GRBs are taken from GCN$^*$, while the redshifts of
GRBs with no redshift measurement are artificially taken to be 1.0
(with superscript of $a$). \label{tbl-1}}
\begin{tabular}{cccccccccc}
\tableline\tableline
GRB Name & $E_{{\rm iso},51}$ & $z$ & $\epsilon_{e,f}$ & $\epsilon_{B,r}$ &
$p_r$ & $M(10^6\,{\rm s})/M_{\odot}$ & $\beta_{\rm max}$ & $r_{m, \rm min}$ (km) \\
\tableline
051016B & 1 & 0.94 & 0.001 & 0.5 & 2.05 & 1.55 & 0.083 & 17.8\\
060109 & 1 & 1.0$^a$ & 0.01 & 0.001 & 2.5 & 1.80 & 0.049 & 23.5\\
060510A & 1 & 1.0$^a$ & 0.01 & 0.5 & 2.1 & 2.43 & 0.084 & 17.7\\
061121 & 10 & 1.31 & 0.1 & 0.5 & 2.3 & 2.07 & 0.12 & 15.5\\
070103 & 0.1 & 1.0$^a$ & 0.01 & 0.5 & 2.05 & 1.42 & 0.097 & 16.7\\
070714B & 1 & 0.92 & 0.01 & $0.0006$ & 2.5 & 1.79 & 0.11 & 17.8\\
080229A & 10 & 1.0$^a$ & 0.1 & 0.4 & 2.2 & 2.30 & 0.13 & 15.0\\
080310 & 1 & 2.43 & 0.1 & 0.002 & 2.2 & 1.92 & 0.11 & 17.5\\
091029 & 1 & 2.75 & 0.1 & 0.4 & 2.2 & 1.66 & 0.073 & 18.6\\
110213A & 10 & 1.46 & 0.1 & 0.005 & 2.2 & 2.32 & 0.13 & 15.4\\
120118B & 1 & 1.0$^a$ & 0.01 & $0.0001$ & 2.5 & 2.29 & 0.053 & 20.7\\
120404A & 10 & 2.87 & 0.1 & 0.002 & 2.2 & 2.07 & 0.083 & 17.8\\
\tableline
\end{tabular}
\end{center}
\begin{list}{}
\item[$^*$]: GRB 051016B (Soderberg et al. 2005), GRB 061121 (Bloom et al. 2006),
GRB 070714B (Graham et al. 2007), GRB 080310 (Prochaska et al. 2008), GRB 091029
(Chornock et al. 2009), GRB 110213A (Milne et al. 2011), GRB 120404A (Cucchiara et al. 2012)\\
\end{list}
\end{table}

\begin{figure}
\plotone{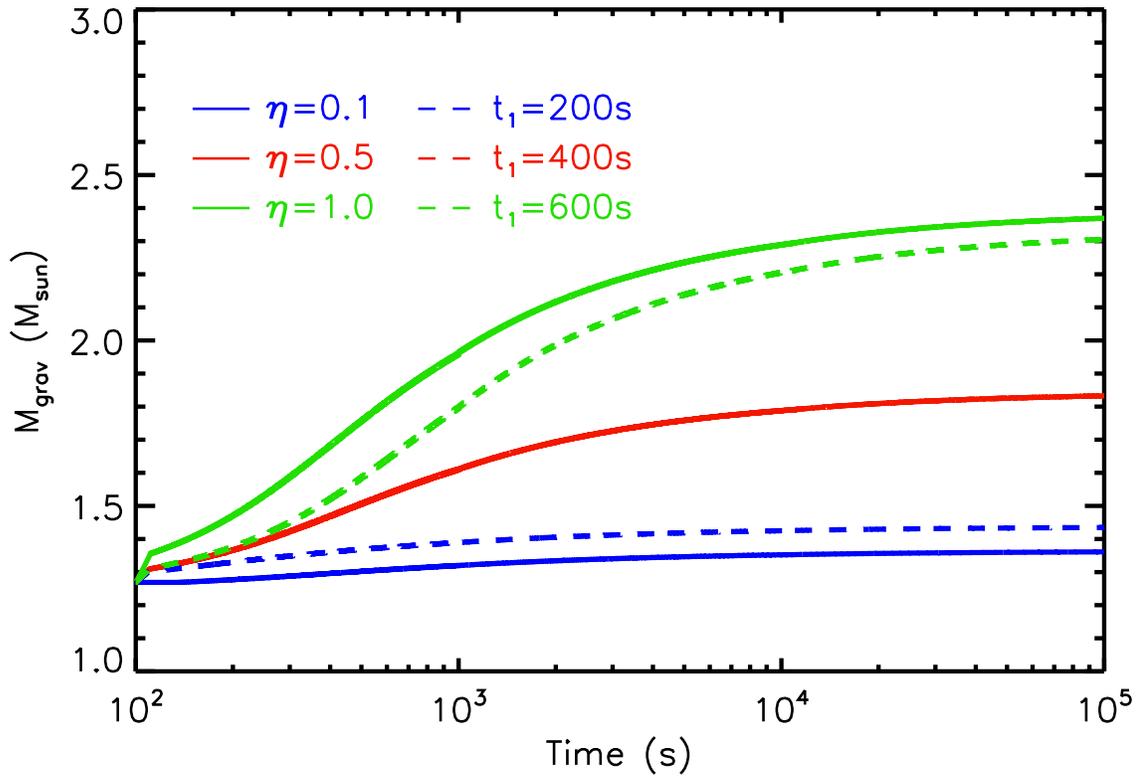} \caption{Gravitational mass evolution of a pulsar
for different values of $\eta$ ({\em solid lines}) and $t_1$ ({\em
dashed lines}). The initial baryonic mass of the pulsar is taken to
be $M_0=1.4M_{\odot}$. \label{fig1}}
\end{figure}

\begin{figure}
\plotone{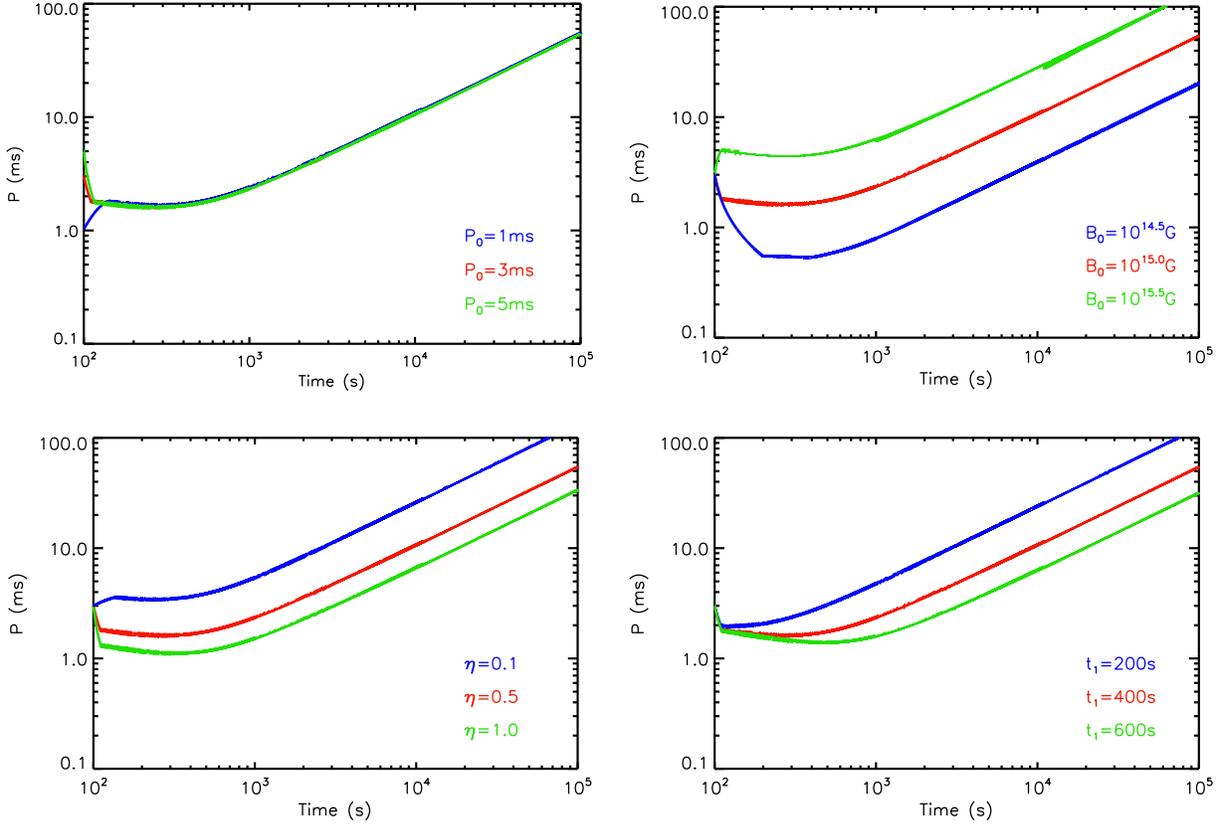} \caption{Spin evolution of a pulsar with time for
different parameters. The initial baryonic mass and radius of the
pulsar are taken to be $M_0=1.4M_{\odot}$ and $R_s=12\,$km
respectively. The benchmark values of the other parameters such as
the initial rotation period $P_0$, surface magnetic field strength
$B_0$, dimensionless accretion rate $\eta$ and transition time of
accretion modes $t_1$ are taken as follows: $P_0=3\,$ms,
$B_0=10^{15}\,$G, $\eta=0.5$, and $t_1=400\,$s. In each panel, we
also plot spin evolution if one of the four parameters is changed
while the other three parameters are fixed to be the benchmark
values. \label{fig2}}
\end{figure}

\begin{figure}
\plotone{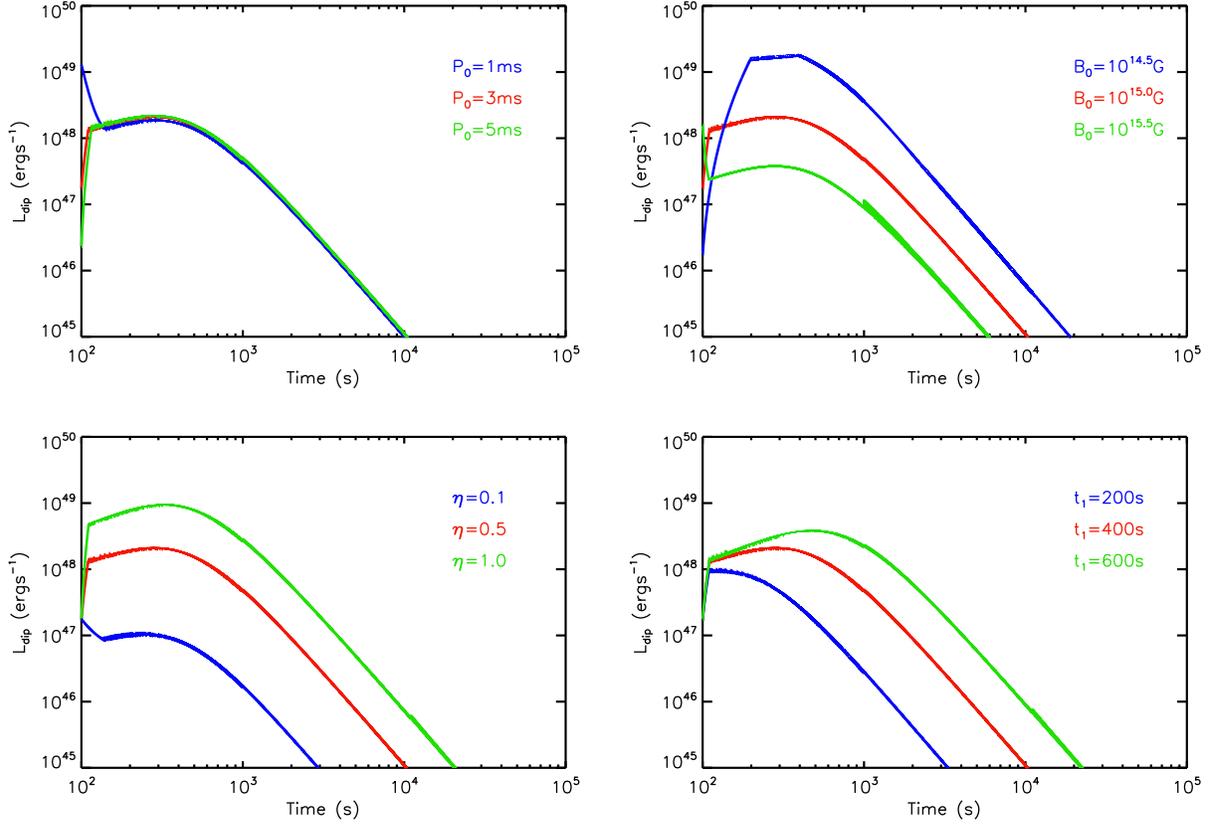} \caption{The same as Fig.~\ref{fig2} but for
evolution of the magnetic-dipole-radiation luminosity $L_{\rm dip}$
with time. \label{fig3}}
\end{figure}

\begin{figure}
\plotone{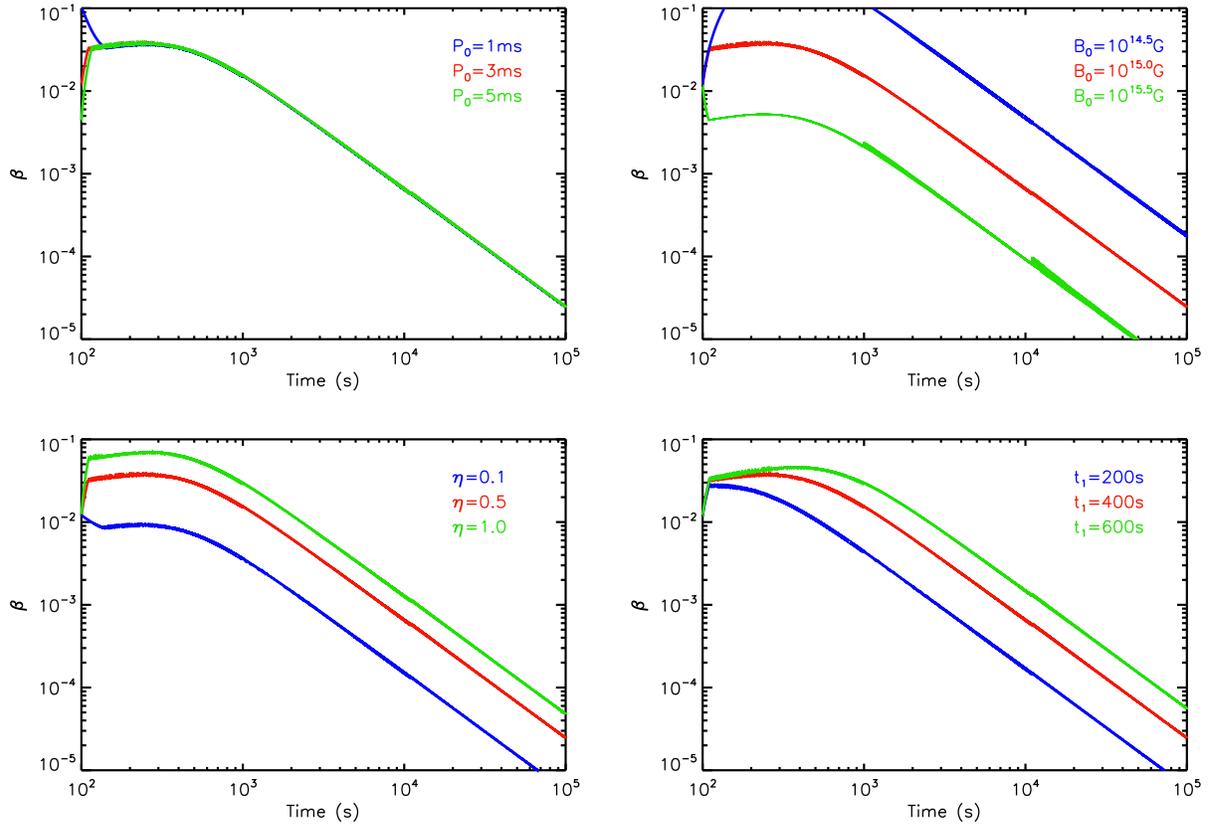} \caption{The same as Fig.~\ref{fig2} but for
evolution of the pulsar's rotation parameter $\beta$ with
time.\label{fig4}}
\end{figure}

\begin{figure}
\plotone{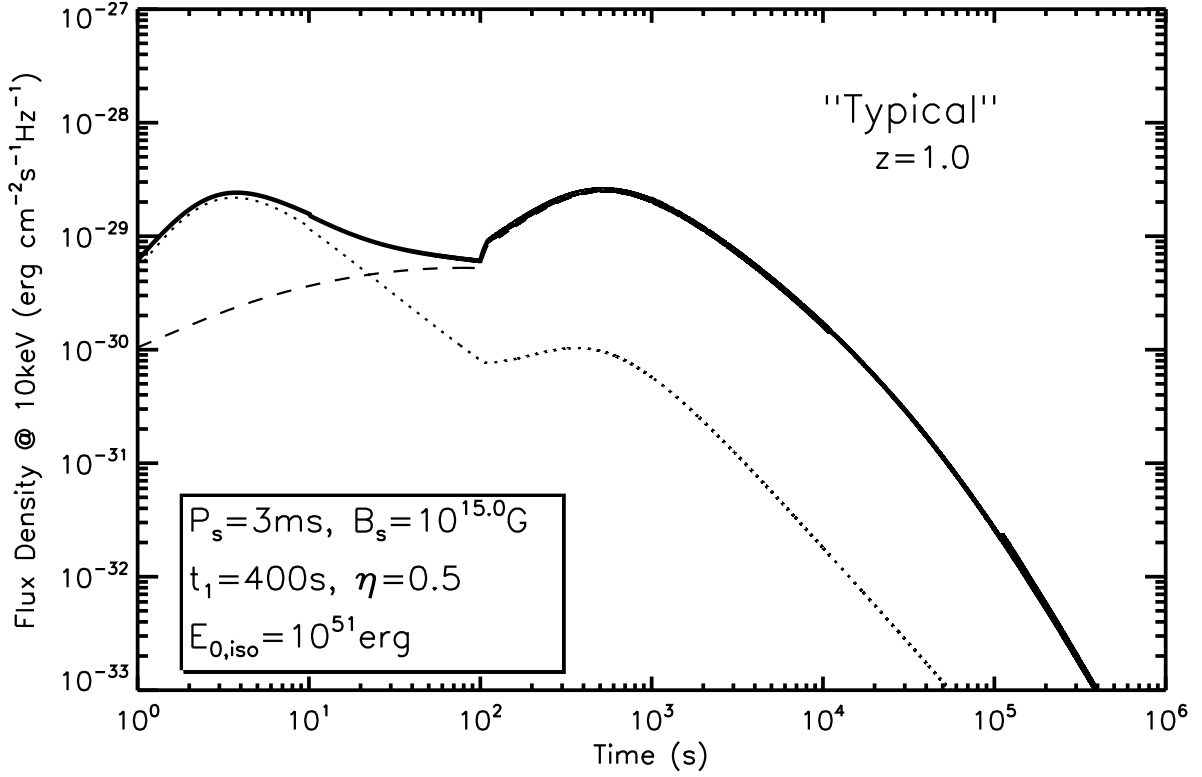} \caption{A typical light curve in our model. The
dashed line is the contribution from the reverse shock emission in a
relativistic wind bubble for the shock parameters
$\epsilon_{B,r}=0.1$, $\epsilon_{e,r}=1-\epsilon_{B,r}=0.9$, and
$p_r=2.5$, while the dotted line is the contribution from the
forward shock emission in the ISM for the shock parameters
$\epsilon_{B,f}=\epsilon_{e,f}=0.1$, and $p_f=2.2$. The initial bulk
Lorentz factor of the GRB fireball is taken to be $\gamma_0=300$ and
that of the pulsar wind of $e^{\pm}$ pairs is $\gamma_{\rm w}=10^6$.
The thick solid line is the sum of the two components. The other
parameters are labeled in the figure.\label{fig5}}
\end{figure}

\begin{figure}
\plotone{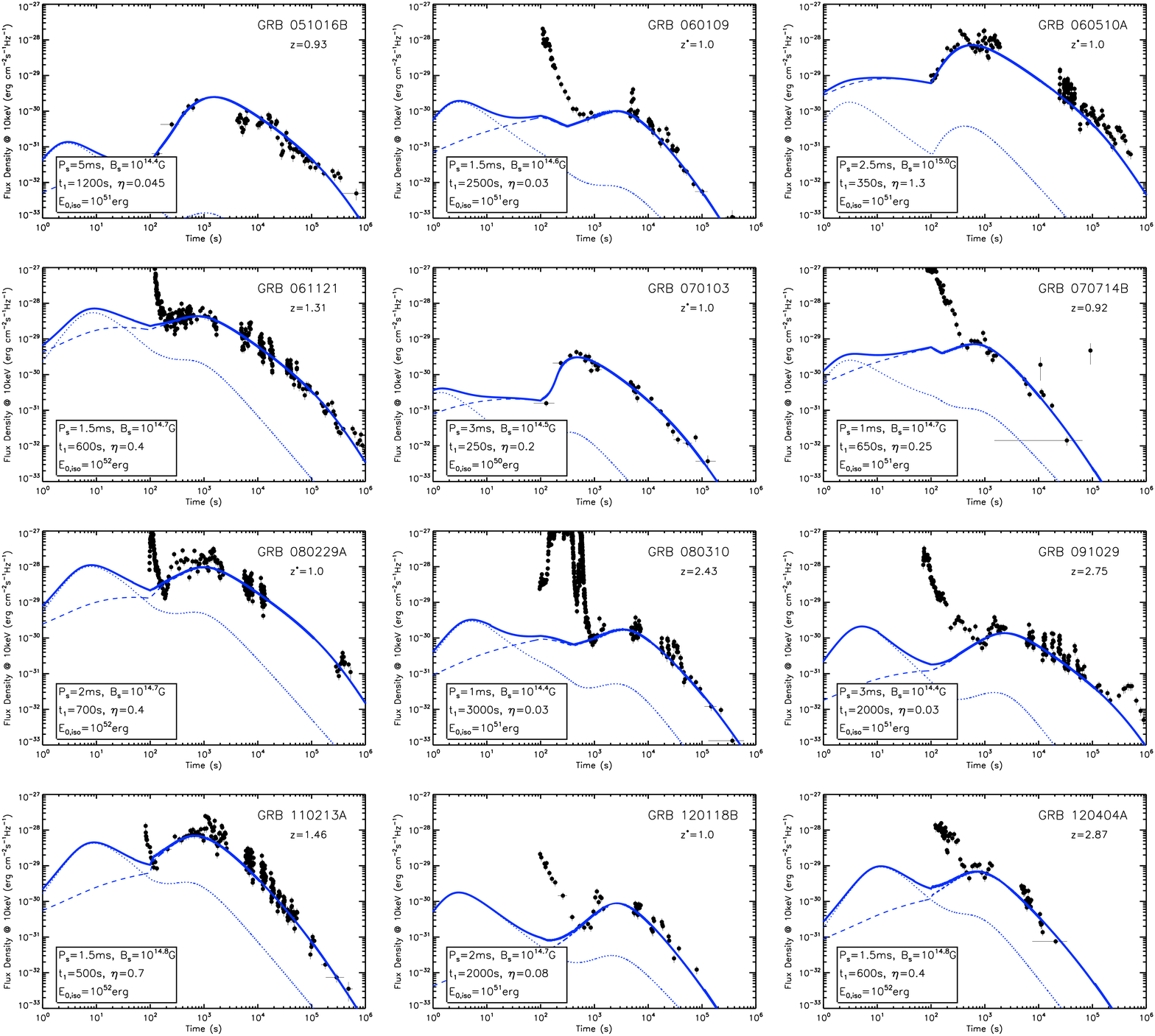} \caption{Fitting to the light curves of 12 X-ray
afterglows with early-time significant brightening in our model. The
black filled circles with error bars are Swift-XRT data. The blue
dashed lines are contributions from reverse shocks, while the blue
dotted lines represent contributions from forward shocks. The thick
blue solid lines are the total fluxes of the two components. The
shock parameters can be found in Table 1. \label{fig6}}
\end{figure}

\end{document}